\documentclass[aps,pre,twocolumn,showpacs,superscriptaddress]{revtex4}

\usepackage{graphicx}
\usepackage{amsmath}
\usepackage{siunitx}

\newcommand*{\vect}[1]{\mathbf{#1}}
\newcommand*{\abs}[1]{\left|#1\right|}
\newcommand*{\avg}[1]{\left<#1\right>}
\newcommand*{\lp}{\ell_{\rm P}}

\newcommand*{\fig}[1]{Fig.~\ref{Fig:#1}}
\newcommand*{\subfig}[2]{Fig.~\ref{Fig:#1}(#2)}
\newcommand*{\eq}[1]{Eq.~(\ref{Eq:#1})}

\begin{document}

\title{Orientational correlations in confined DNA} 

\author{E. Werner\footnote{E. Werner and F. Persson contributed equally to this work.}}
\author{F. Persson$^*$}
\affiliation{Department of Physics, University of Gothenburg, Sweden}
\author{F. Westerlund}
\affiliation{ Department of Chemical and Biological Engineering, Chalmers University of Technology, Sweden}
\author{J. O. Tegenfeldt}
\affiliation{Department of Physics, University of Gothenburg, Sweden}
\affiliation{Department of Physics, Division of Solid State Physics, Lund University}
\author{B. Mehlig}
\affiliation{Department of Physics, University of Gothenburg, Sweden}
\date{\today}

\begin{abstract}
We study how the orientational correlations of DNA confined to nanochannels depend on the 
channel diameter $D$ by
means of Monte Carlo simulations and a mean-field theory. 
This theory describes DNA conformations in the experimentally
relevant regime where the Flory-de Gennes theory does not apply.
We show how local correlations determine the dependence of the end-to-end distance of the DNA
molecule upon $D$. 
Tapered nanochannels provide the necessary resolution in $D$ to study experimentally 
how the extension of confined DNA molecules depends upon $D$. Our experimental and theoretical results are in qualitative agreement.
\end{abstract}

\pacs{82.35.Lr, 87.14.gk, 87.15.A-}

\maketitle
The conformations of biopolymers in living systems are often affected by confinement \cite{marenduzzo2010,zhou2008}. 
Examples include actin and its analogs in the gel-like cytoplasm \cite{koster2009}, DNA segregation in bacterial chromosomes \cite{jun2006}, and the dense DNA packing in eukaryotic chromosomes \cite{wolffe1998}, in bacterial spores \cite{errington1993}, and in viral capsids and tail tubes \cite{speir2012,inamdar2006,lof2007}. Restrictions of the available conformations fundamentally influence function, for example in the case of DNA condensation \cite{baumann2000,zhang2012}. 

\begin{figure}
  \includegraphics[width=8.5cm]{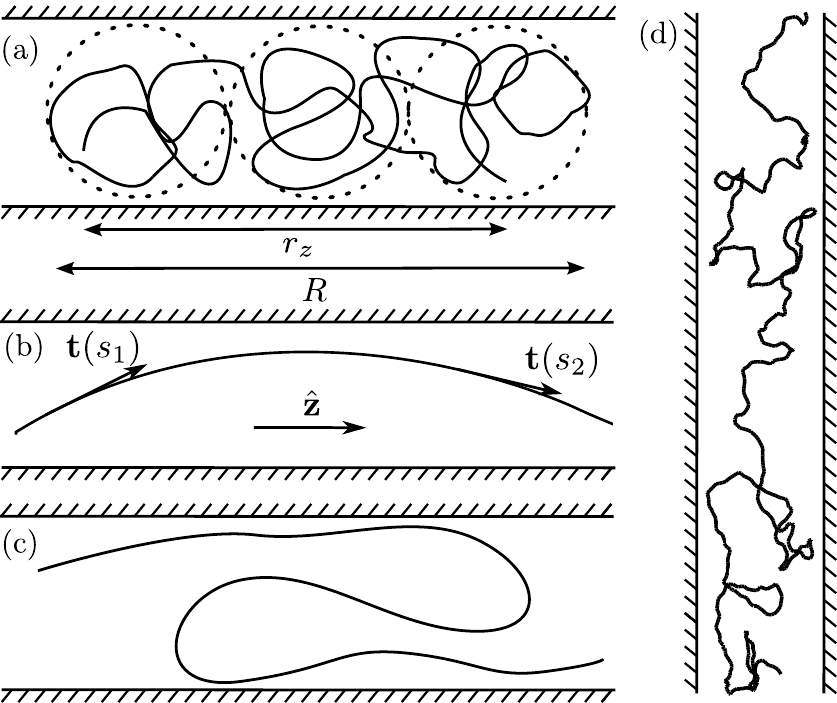}
  \caption{Schematic illustration of different regimes for confined DNA. (a) The de Gennes regime ($D\gg \lp$). (b) The Odijk regime ($D\ll \lp$). (c) The Odijk regime with hairpin formation ($D\lesssim \lp$). (d) A snapshot 
from Monte Carlo simulations (described in the text) 
of a semiflexible chain confined to a square channel with diameter $D=4.45\lp$.} 
  \label{Fig:illustration} 
\end{figure}

Single DNA molecules confined in nanofluidic channels are a powerful model system for studying the physics of confined 
biopolymers in well-controlled environments \cite{PerssonReview2010,LevyReview2010,persson2009EPA}. Understanding the behavior of this model system is thus a first step toward understanding the effects of confinement on polymers in more complex biological systems. 
The principal difficulty is that since the persistence length of DNA ($\lp\approx \SI{50}{\nano\meter}$) is below the diffraction limit for visible light, its microscopic configurations are not directly observable in the fluorescence microscope. To infer the statistics of local conformations
it is therefore crucial to understand theoretically how such local conformations determine large-scale observables,
such as, for example, the extension $R$ of the confined DNA molecule [\subfig{illustration}{a}].

Two generally accepted theories exist for how the extension of a DNA molecule varies as a function of channel diameter $D$.
De Gennes' scaling theory \cite{deGennesBook}, valid for wide channels ($D\gg \lp$), predicts that $R \propto D^{-2/3}$. 
It is based on the notion that a DNA molecule can be divided into a sequence of ``blobs'' [\subfig{illustration}{a}], 
and that the DNA within a blob follows Flory scaling \cite{Flory1953}. Odijk's theory \cite{Odijk1983}, by contrast, 
describes the conformations of very strongly confined DNA ($D\ll \lp$) as almost stiff segments deflecting from the channel walls [\subfig{illustration}{b}].

Between these asymptotic limits, the local mechanisms determining the extension of the molecule are not understood, yet this regime is where most experiments are conducted. 
This third regime spans at least an order of magnitude in $D$,
and cannot be understood in terms of either asymptotic picture. 
Several authors have attempted to understand the dependence of the extension upon the channel diameter in this regime in terms of power-law relations of the form $R\propto D^{-\alpha}$ \cite{reisner2005,cifra2009chain,wang2011}.
Instead we demonstrate that this third regime can be successfully analyzed in terms of the orientational correlation function $C_z(s_1,s_2)=\avg{t^z(s_1) t^z(s_2)}$. 
Here $t^z(s)$ is the $z$-component of the unit tangent vector at contour distance $s$ from the beginning of the chain [\subfig{illustration}{b}]. 
Angular brackets denote a time average.
A surprisingly simple picture emerges: the orientational correlations of confined DNA exhibit three distinct behaviors that can easily be separated.
At short separations the correlation function decays exponentially. We show that the parameters of this decay depend upon $D$, 
and discuss the mechanisms behind this dependence. At larger separations, self-avoidance dominates the correlations, giving rise to a plateau in $C_z$ which can be understood in terms of a mean-field theory. 
This theory applies in the experimentally relevant range where the DNA conformations cannot be described
in terms of statistically independent blobs [\subfig{illustration}{d}]. Moreover, the theory shows that
the dependence of the extension upon $D$ is not of power-law form for the range of
parameters that are easily accessible 
in experiments and numerical simulations.
Despite the fact that the mean-field theory is not based on the notion of blobs, it predicts that $R \propto D^{-2/3}$, consistent with de Gennes scaling, in the asymptotic limit 
of very long DNA in wide channels. 
However, in experiments and simulations, the contour length is usually not long enough for the asymptotic regime to be reached.
We show that in this case end effects (quantified by the decay of $C_z$ at very large separations) substantially modify the fluctuations of $R$.
We test our model by Monte Carlo simulations of a semiflexible, self-avoiding polymer, and by experimentally measuring the dependence of the extension $R$ of a confined DNA molecule upon $D$ in tapered nanochannels (where the channel diameter $D$ varies gradually along the channel) \cite{persson2009CS}, that provide the necessary $D$ resolution to test the theoretical predictions. We find qualitative agreement between the theoretical and experimental results.
2
A DNA molecule in solution is commonly described as a wormlike chain of persistence length $\lp\approx\SI{50}{\nano\meter}$ and contour length $L$, where different pieces interact through a screened repulsive electrostatic potential. In order to simplify the theory, we instead consider a semiflexible chain of $N$ spherical monomers of diameter $a$, with a bending potential $U_i/(k_B T)=-\kappa (\vect{t}_{i}\cdot\vect{t}_{i-1})$.
Here $\vect{a}_{i}=a\vect{t}_{i}$ points from the center of monomer $i$ to the center of monomer $i+1$ and $\vect{t}_i$ is the unit tangent vector. $\kappa$ is a dimensionless measure of the stiffness. Defining $\lp=\kappa a$, the wormlike chain can be recovered in the limit $\kappa\to \infty$, $a\to 0$, $\lp=\kappa a={\rm const.}$ Since the electrostatic interaction is short-ranged, it can be approximated by a hard-core potential with an 
effective width $w_{\rm eff}$ which depends on the ionic strength of the solution. Let us constrain the centers of the monomers to a square channel of width and height $D$, extending along the $z$-direction. The end-to-end distance of the chain is given by $r_z=a\sum_i t_i^z$, where $t_i^z=\vect{t}_i\cdot \hat{\vect{z}}$ (\fig{illustration}). We characterize the orientational statistics of the confined DNA molecule by the correlation function $C_z(i,j)=\avg{t_i^z t_j^z}$. Since the confinement breaks the symmetry, the $z$-component of the tangent behaves very differently from the $x$- and $y$-components. 
The relevant correlation function is thus $C_z(i,j)$, as opposed to $\avg{\vect{t}_i\cdot \vect{t}_j}$. The latter function 
has been studied for free DNA \cite{schafer1999,wittmer2004,hsu2010}, for strongly confined DNA ($D<\lp$) \cite{wagner2007,cifra2010}, and for strongly confined actin filaments \cite{choi2005, koster2008}.

\begin{figure}
  \includegraphics[width=8.5cm]{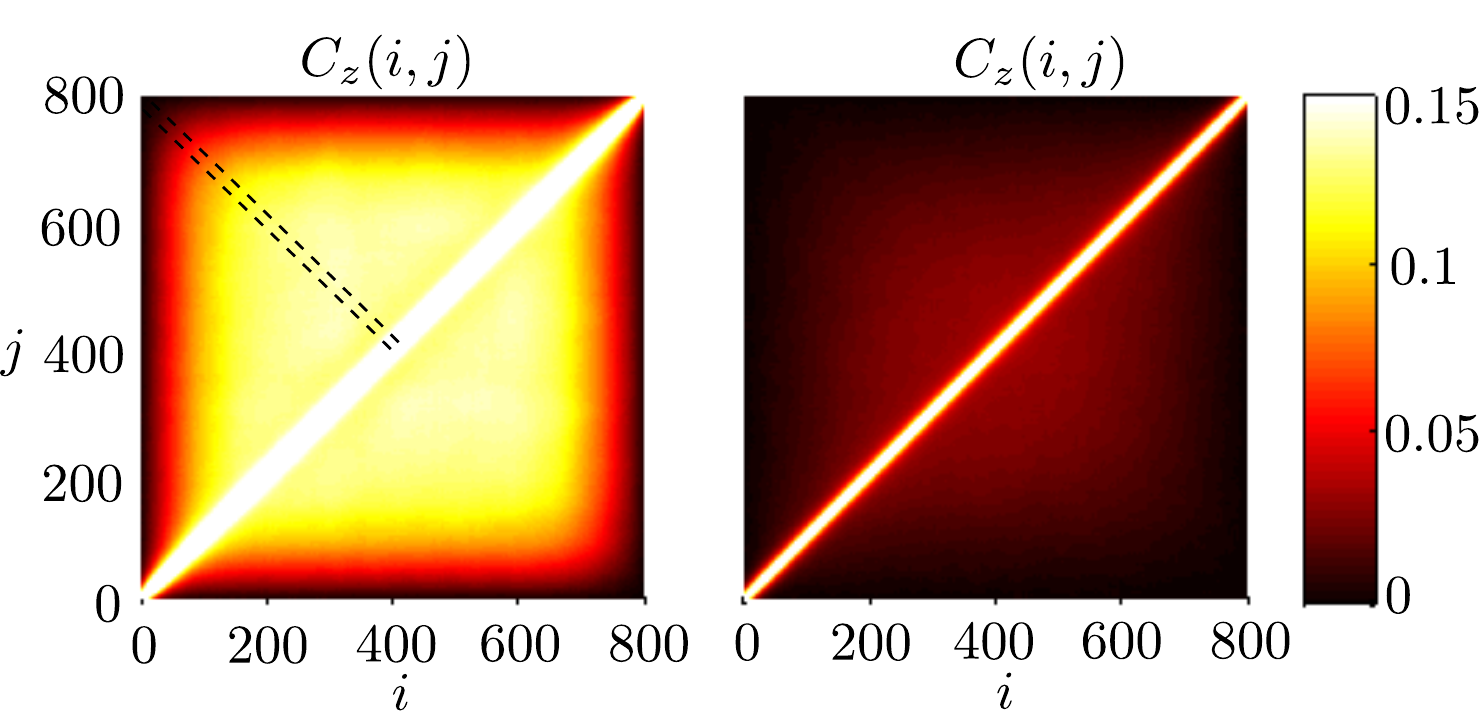}
  \caption{(Color online) Correlation functions $C_z(i,j)$ from Monte Carlo simulations, 
constrained to square channels. Left: $D=2\,\lp\equiv 2 \kappa a$. Right: $D=4.45\lp$. Regions where
$C_z > 0.15$ are white. Dashed lines mark the section that is shown in Figs. \ref{Fig:multiplot}(a) and \ref{Fig:multiplot}(b).}
  \label{Fig:surfs}
\end{figure}

\begin{figure}
  \includegraphics[width=8.3cm]{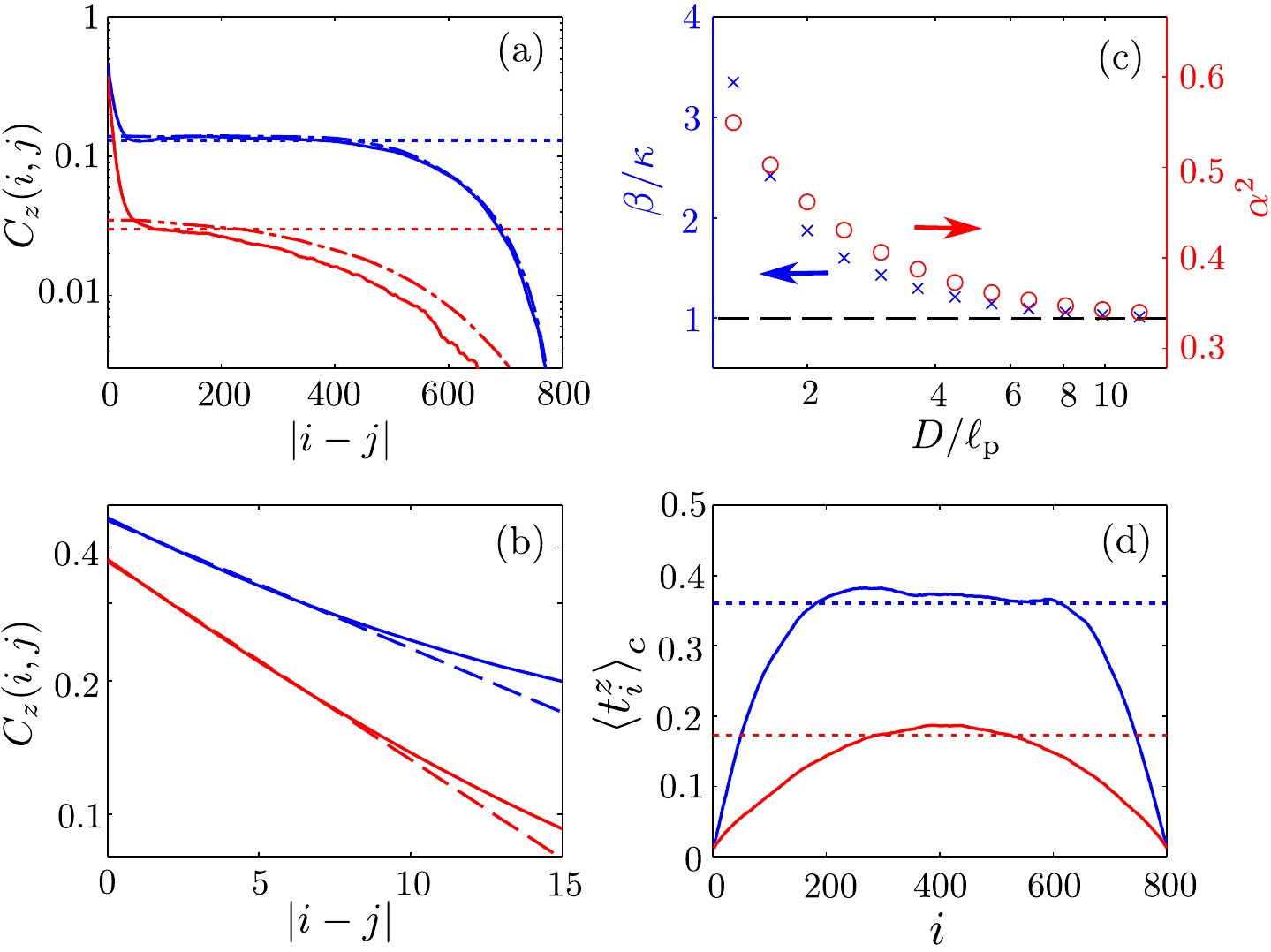}
  \caption{(Color online) (a, b) Sections of the correlation function $C_z(i,j)$. Blue (upper) line: $D=2\,\lp$. Red (lower) line: $D=4.45\lp$. $C_z$ is averaged over all values with $\abs{i+j-N}<10$, i.e., between the dashed lines in \fig{surfs}. Dotted lines: estimates of the plateau level $C_z=\avg{t^z}_{\rm c}^2$ from \eq{r_theory}. Dashed lines: exponential fits for $\abs{i-j}\le \kappa$. Dash-dotted lines: $\avg{t_i^z}_{\rm c} \avg{t_j^z}_{\rm c}$ (hardly distinguishable from solid line for $D=2\lp$). (c) Estimates of the decay parameters $\alpha^2$ (red circles), $\beta$ (blue crosses), as a function of $D/\lp$. (d) $\avg{t_i^z}_{\rm c}$ [with $\alpha(D)$ and $\beta(D)$ taken
 from (c)] as a function of $i$ for $D=2\,\lp$ (blue, upper line) and $D=4.45\lp$ (red, lower line). Dotted lines: estimates of $\avg{t^z}_{\rm c}$ from \eq{r_theory}. }
  \label{Fig:multiplot}
\end{figure}

In our simulations, non-neighboring monomers interact by a hard-core potential with effective width $w_{\rm eff}=a$. We use $N=800$, $\kappa=8$ and $a=1$, corresponding to a DNA molecule with $w_{\rm eff}\approx \SI{6}{\nano\meter}$ and contour length $L\approx\SI{5}{\micro\meter}$ (assuming $\lp\approx \SI{50}{\nano\meter}$). The model and the parameters are similar to the ones in Ref.~\cite{wang2011}. The simulations implement the Metropolis algorithm, with crankshaft trial moves \cite{yoon1995}. The resulting orientational correlation functions $C_z(i,j)$ are shown in \fig{surfs} for two different values of the channel size $D$, and
\subfig{multiplot}{a,b} shows sections of the correlation functions corresponding to the region between the dashed lines in \subfig{surfs}{left}.


$C_z(i,j)$ is seen to exhibit three distinct behaviors we now describe in turn. First, when $\abs{i-j}\lesssim \kappa$, the effect of self-avoidance is expected to be negligible 
compared to the effect of stiffness. For an unconfined chain ($D\gg\lp$) the correlation function decays as $C_z(i,j)=\alpha^2\exp\{-|i-j|/\beta\}$ in this region, with $\alpha^2=1/3$, $\beta=\kappa$. The results of our simulations show that the decay remains approximately exponential for smaller channels ($D\gtrsim\lp$), \subfig{multiplot}{b}. We have fitted $\alpha$ and $\beta$, and find that both parameters depend upon $D$ [\subfig{multiplot}{c}]. We see that $\alpha$ increases as $D$ decreases, reflecting a tendency of the segments to align with the channel direction. This observation is consistent with recent experimental findings \cite{persson2009EPA}.
The parameter $\beta$ quantifies the initial decay of orientational correlations of confined DNA. We expect that for $D\lesssim \lp$, the parameter $\beta$ is determined by the probability of hairpin formation. The corresponding free energy was calculated in Ref.~\cite{odijk2006}. However, a theory for $\beta$ in this regime is lacking, and $\beta$ must be determined by simulations. This also applies in the case of the wider channels we consider here.

Second, for $\abs{i-j}\gg \beta$, the tangent vectors at $i$ and $j$ are independently oriented, except for the fact that self-avoidance swells the chain. Thus, $C_z(i,j)=\avg{t_i^z}_{\rm c}\avg{t_j^z}_{\rm c}\,$, where the subscript defines an average conditional on $r_z>0$. Whereas $\avg{t_i^z}=0$ due to the $z$-symmetry of the problem, $\avg{t_i^z}_{\rm c}$ takes into account the fact that strong confinement breaks the $z$-symmetry: $r_z$ is rarely close to zero for a long chain in a thin channel. Figure~\ref{Fig:multiplot}(a) shows that this factorization of $C_z$ works very well for thin channels.
The question is now how $\avg{t_i^z}_{\rm c}$ depends on $i$. For monomers far from the ends,
$\avg{t_i^z}_{\rm c}$ is expected to be independent of $i$ (in this case we write $\avg{t_i^z}_{\rm c}=\avg{t^z}_{\rm c}$); compare \subfig{multiplot}{d}. As a consequence, a plateau in $C_z(i,j)$ develops, 
clearly seen in Figs.~2 and 3(a).
We show below how the bias $\avg{t^z}_{\rm c}$ (and thus the level of the plateau) can be estimated by a mean-field argument. 
Fig.~\ref{Fig:surfs}(right) shows results for a channel that is so wide that a plateau does not clearly develop for a chain of this length. This fact is also apparent in \subfig{multiplot}{a}, solid red line.

Third, the bias $\avg{t_i^z}_{\rm c} $ is expected to be smaller close to the ends, as the chain is more flexible there. This is clearly seen in Fig.~\ref{Fig:multiplot}(d) and gives rise to a further decay of $C_z(i,j)$, seen in Fig.~\ref{Fig:surfs} and Fig.~\ref{Fig:multiplot}(a). We note that such end decay has been observed in other contexts too \cite{schafer1999,cifra2010,hsu2010}. However, a quantitative theory for the onset and shape of the decay for confined DNA molecules is lacking. 

We now discuss the implications of our results for $C_z$ for the extension of a confined DNA molecule. Let us first consider the limit of very long chains ($N\rightarrow \infty$). Neither our study nor other experimental and simulation studies achieve this regime, but it is nevertheless instructive to consider. In this limit, fluctuations are negligible, and the extension is almost equal to the end-to-end distance, which in turn is determined by the correlation function. Thus, $R^2\approx\avg{r_z^2}=a^2\sum_{i,j} C_z(i,j)$. Since the contributions to the sum from the first and third regions scale linearly with $N$, whereas that from the second region grows as $N^2$, the extension is determined by the bias, $R^2=N^2a^2 \avg{t^z}_{\rm c}^2$, as $N\rightarrow\infty$. We note that in this limit, the extension $R$ scales linearly with $L=N a$, as it must when self-avoidance dominates the extension. 

We now demonstrate how the bias $\avg{t^z}_{\rm c}$ (and thus $R$ for $N\to \infty$) can be estimated by a mean-field argument. One way to obtain the correct average $\avg{\cdots}_{\rm c}$ for a self-avoiding polymer is by first generating all configurations of the corresponding ideal polymer (i.e., with spatial overlaps allowed), and then remove all configurations where two or more monomers overlap. If ${P}_{\rm ideal}(r_z)$ is the probability distribution of $r_z$ for the ideal chain, the distribution for the self-avoiding chain is given by $P(r_z)\propto P_{\rm ideal}(r_z)A(r_z)$, where $A(r_z)$ is the fraction of ideal configurations with end-to-end distance $r_z$ that are free of overlaps. Estimating the functions $P_{\rm ideal}(r_z)$ and $A(r_z)$ leads to an approximate expression for $P(r_z)$. Let us start with $P_{\rm ideal}$. Unless the channel is very thin ($D\ll \lp$), the correlation function $C_z^{\rm ideal}(i,j)$ decays rapidly to zero for $\abs{i-j} > \kappa$. Assuming that the correlation 
decays exponentially as $C_z^{\rm ideal}(i,j)=\alpha^2\exp\{-|i-j|/\beta\}$, where $1\ll\beta\ll N$, the distribution $P_{\rm ideal}(r_z)$ is a Gaussian function, with zero mean and variance $\avg{r_z^2} \approx N a^2\alpha^2 2\beta$. We now estimate $A(r_z)$ by a mean-field argument, similar to the one used in deriving the Flory expression for the extension of a free self-avoiding chain \cite{Flory1953,deGennesBook}. If we divide our polymer into $N_\beta=N/(2\beta)$ effective monomers of length $2\beta a$ and width $a$, they are essentially independently oriented. If we make the mean-field assumption that these effective monomers are uniformly and independently distributed within the available volume $V=r_z D^2$, the probability for two given monomers to collide is $p=\xi/V$. 
Here $\xi$ is the excluded volume of an effective monomer. We approximate $\xi$ by 
the value for a stiff rod of the same length and width \cite{Onsager1949}: $\xi=({\pi}/{2})a^3(4\beta^2+(\pi+3)\beta+\pi/4)$. Since there are $N_\beta(N_\beta-1)/2\approx N_\beta^2/2$ possible collisions, the probability of no collisions is $ A(r_z)=(1-p)^{N_\beta^2/2}= [1-{\xi}/(r_z D^2)]^{N_\beta^2/2}$. With these expressions for $P_{\rm ideal}$ and $A$, we find $P(r_z)$. Differentiation yields the most probable end-to-end distance, and thus an estimate of the bias:
\begin{equation}
\label{Eq:r_theory} \avg{t^z}_{\rm c}
\approx \left(\frac{\xi(D)\alpha^2(D)}{4 \beta(D) a D^2}\right)^{1/3}\,.
\end{equation}
Note that $\alpha^2$ and $\beta$ (and thus $\xi$) depend on $D$. The prediction of \eq{r_theory} -- with $\alpha(D)$ and $\beta(D)$ taken from \subfig{multiplot}{c} -- is compared to simulation results in Figs. \ref{Fig:multiplot}(a) and \ref{Fig:multiplot}(d). The agreement is surprisingly good considering the shortcomings of the mean-field theory: First, the expression for the excluded volume assumes that the effective monomers are randomly oriented stiff rods, while in reality they have complicated shapes and have a tendency to align with the channel. This assumption overestimates the bias by an unknown factor in the region where $\alpha^2\gg 1/3$. Second, our theory assumes that monomers are uniformly distributed within the channel, whereas in fact monomers are more likely to be found in the center of the channel than near the walls. This assumption underestimates the bias by a factor of order unity, for all values of $D$. 

Equation (\ref{Eq:r_theory}) shows that the dependence of $R$ on $D$ is not of power-law form in the limit $N\to \infty$, except for very wide channels, where $\alpha^2\to 1/3$ and $\beta \to \kappa$ [\subfig{multiplot}{c}]. 
In this limit, \eq{r_theory} gives $R\propto D^{-2/3}$, in agreement with de Gennes scaling. We emphasize that our derivation is not based on the notion that DNA within a blob of size $D$ follows Flory scaling, but
directly takes into account the confinement of the DNA molecule in the channel. The theory is valid for channels so thin that DNA conformations cannot be described
in terms of statistically independent blobs [\subfig{illustration}{d}]. Flory scaling is not relevant to this regime, neither is the question which Flory
exponent should be used ($\nu = 3/5$, $1/2$, or $0.588$) \cite{wang2011,cifra2009channel}.

For finite values of $N$ the situation is yet more complicated, as the relative areas of the three regions
in the correlation function change with $D$. This is a second reason why the dependence of $r_z$ upon $D$ is in general not of power-law form.

\begin{figure}
  \includegraphics[width=8.3cm]{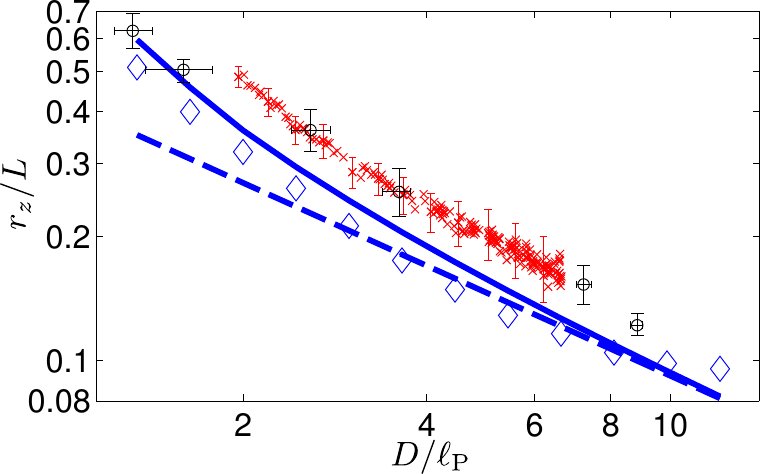}
  \caption{(Color online) DNA extension from simulations and experiments. $\sqrt{\avg{r_z^2}}/L$ from simulations (blue diamonds) and $\avg{t^z}_{\rm c}$ from \eq{r_theory} (solid blue line) compared to de Gennes scaling (dashed blue line). Extension of ${\lambda}$-DNA (48.5 kbp, $L\approx \SI{20}{\micro\meter}$ \cite{PerssonReview2010}) as a function of $D=\sqrt{D_h D_w}$ in \SI{180}{\nano\meter} deep nanofunnels studied at 0.5x TBE (red crosses). The ``error bars'' show the standard deviation of the distribution of extensions, at a given $D$ \cite{sm}. Also shown is experimental data at 0.5x TBE from Fig.~5 of \cite{reisner2005} (black circles).}
  \label{Fig:exp}
\end{figure}

We have verified these conclusions by Monte Carlo simulations. Figure~\ref{Fig:exp}(a) shows the measured end-to-end-distance $\sqrt{\avg{r_z^2}}$ as a function of $D$ 
compared with the scaling law $r_z \propto D^{-2/3}$. Also shown is $\avg{t^z}_{\rm c}$ obtained from Eq.~(\ref{Eq:r_theory}), with $\alpha(D)$ and $\beta(D)$
taken from Fig.~\ref{Fig:multiplot}(c). $\sqrt{\avg{r_z^2}}/L$ and $\avg{t^z}_{\rm c}$ show qualitatively the same behavior, neither of them depending as a power law upon $D$. However, the shape of the curves is clearly different. This is a consequence of the fact that the first and third regions contribute to the extension -- as $D$ increases (at fixed $N$), the end decay starts earlier, thus decreasing the extension faster than predicted by \eq{r_theory}. For even larger values of $D$, the second region vanishes, and the value of $\avg{r_z^2}$ approaches that of free chain.

In order to validate the results from theory and simulations we measured the extension of YOYO-labeled $\lambda$-phage DNA (see Ref.~\cite{PerssonReview2010}). The extension was measured as a function of confinement in tapered nanochannels, with a fixed height ($D_h=\SI{180}{\nano\meter}$) and a gradually increasing width (from $D_w=\SI{50}{\nano\meter}$ to \SI{650}{\nano\meter}). In tapered nanochannels the DNA extension can be measured continuously as the width increases (for experimental details and data for additional ionic strengths we refer the reader to the Supplemental Material \cite{sm}).
For practical reasons the experimental conditions could not be fully reproduced in the simulations as follows: Whereas the simulations were performed in square channels, in the experiment the aspect ratio varies with confinement. Experiment and simulations also differ in that the experimental contour length is approximately 4 times longer, and that the extension $R$ is not identical to the end-to-end distance $r_z$. Furthermore, there is no well-established consensus in the literature on the effects of intercalators on the persistence length of DNA \cite{murade2009,murade2010,vladescu2007,gunther2010} nor is the effect on the effective width known. 
Finally, the interaction between the DNA and the channel walls could lead to a lower effective 
diameter. For these reasons it is hard to compare simulations and experiments quantitatively, and we restrict ourselves 
to noting that they are in qualitative agreement, as shown in \fig{exp}. In particular, the extension curve is always steeper than predicted by de Gennes scaling, and does not obey a power law. Both observations are in accordance with Eq.~(\ref{Eq:r_theory}) and are also consistent with the observation that the end-decay regions grow with increasing channel size.	

The results summarized in this paper pose many new questions. First, our theory includes the scaling prediction of de Gennes for $D\gg \lp$, but what does the transition to the Odijk regime look like? 
Second, how do our results cross over to the known orientational correlations of unconfined DNA \cite{hsu2010}?
Third, it is necessary to understand how the increased fluctuations near the 
ends of the molecule depend on the channel dimension $D$ and the properties of the DNA molecule. Fourth, we have seen
that the parameters characterizing the initial exponential decay of the correlation
function depend upon $D$. This effect is not quantitatively understood. Fifth, we have analyzed a simplified model disregarding possible effects of electrostatic interactions between the molecule
and the walls. These effects are likely to be of importance in the experiments we have discussed, and must be investigated. Sixth, what does the nonuniform monomer distribution of a confined polymer imply for the extension?
Last but not least it is now experimentally possible to study DNA below the diffraction limit of light \cite{persson2011}. 
In the future, these techniques may allow us to directly observe the microscopic confirmations of confined DNA.

\begin{acknowledgments}
This work was supported by Vetenskapsr\aa{}det (BM, JT), the G\"oran Gustafsson Foundation for Research in Natural Sciences and Medicine (BM), the Knut and Alice Wallenberg Foundation (FW), the Swedish Foundation for Strategic Research (FW), 
and the Seventh Framework Program [FP7/2007-2013] under grant agreement number [HEALTH-F4-2008-201418] entitled READNA.
\end{acknowledgments}

\end{document}


\title{{\large \rm Supplemental material for\\} Orientational correlations in confined DNA} 

\author{E. Werner\footnote{E. Werner and F. Persson contributed equally to this work.}}
\author{F. Persson$^*$}
\affiliation{Department of Physics, University of Gothenburg, Sweden}
\author{F. Westerlund}
\affiliation{ Department of Chemical and Biological Engineering, Chalmers University of Technology, Sweden}
\author{J. O. Tegenfeldt}
\affiliation{Department of Physics, University of Gothenburg, Sweden}
\affiliation{Department of Physics, Division of Solid State Physics, Lund University}
\author{B. Mehlig}
\affiliation{Department of Physics, University of Gothenburg, Sweden}
\date{\today}

\maketitle

\section{Experimental details}
\noindent{\bf Imaging.} DNA imaging was made using an epifluorescence video microscopy system consisting of a Nikon Eclipse TE2000-U inverted microscope, 60x water immersion (NA 1.0, Nikon) objective (with a 1.5x additional magnification lens for a total magnification of 90x) coupled to a back-illuminated EMCCD camera from Photometrics, Cascade II 512.\\ 

\noindent{\bf Chip design.} The measurements were performed in a nanofluidic chip. The imaged part of the chip consists of channels with a constant depth of $D_{\rm h}=\SI{180}{\nano\meter}$, and a width $D_{\rm w}$ that varies continuously from \SI{50}{\nano\meter} to \SI{650}{\nano\meter} over a length of \SI{480}{\micro\meter}. The channels were defined by electron beam and UV lithography followed by CF4/CHF3 based reactive ion etching. Finally a thin layer (\SI{50}{\nano\meter}) of dry thermal oxide was grown to render the surfaces hydrophilic. The devices were sealed with a \SI{500}{\micro\meter} thick borosilicate glass lid.\\

\begin{figure}
\includegraphics[width=12cm]{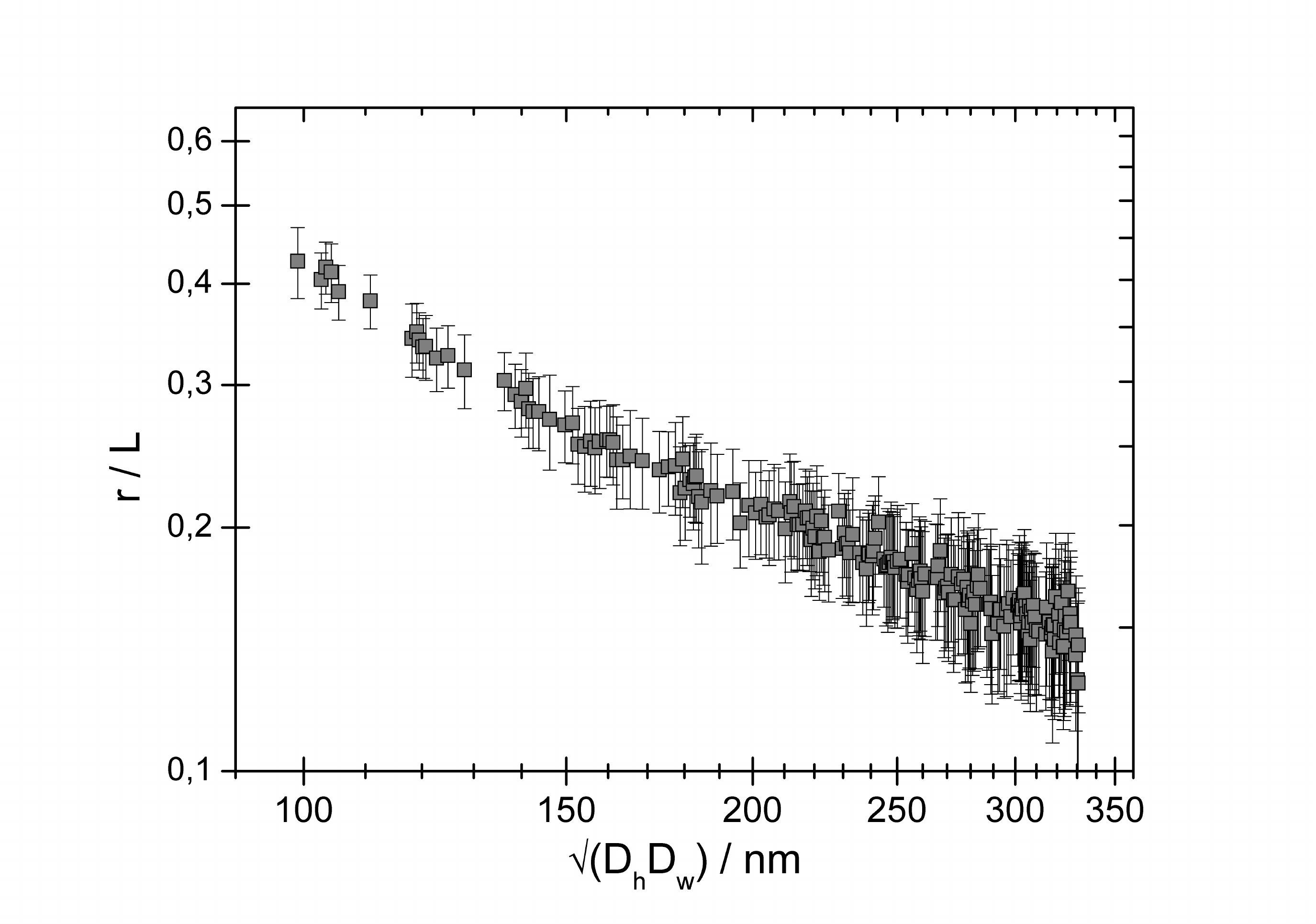}
\caption{The extension $r$ of $\lambda$-DNA (contour length $L\approx \SI{20}{\nano\meter}$) in 1x TBE buffer, as a function of the average diameter $D$ (geometric average of width and depth), in \SI{180}{\nano\meter} deep
nanochannels. The ``error bars'' denote the standard deviation of the measured extension at a given $D$.
} 
\label{Fig:1} 
\end{figure}

\begin{figure}
\includegraphics[width=12cm]{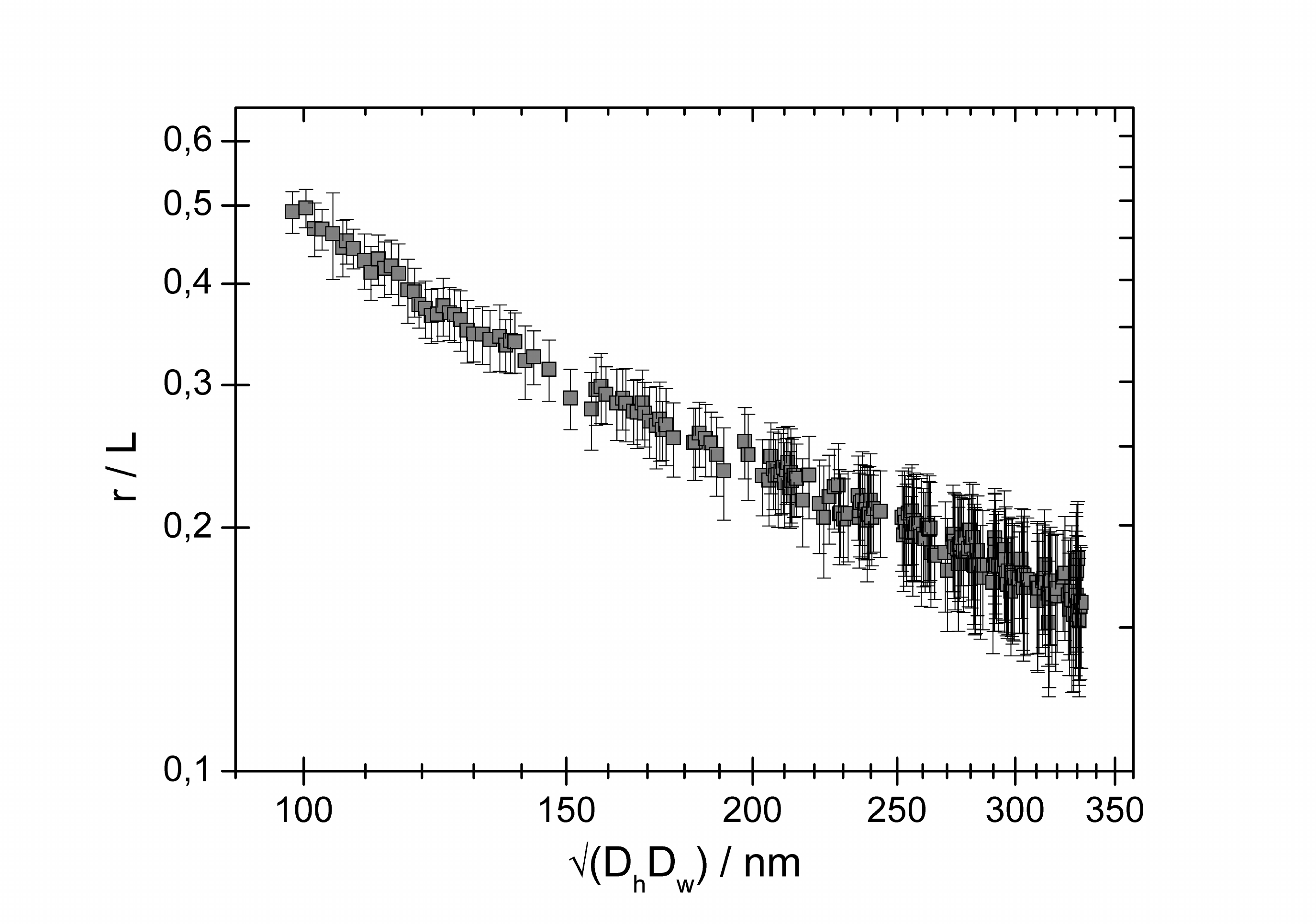}
\caption{The extension $r$ of $\lambda$-DNA (contour length $L\approx \SI{20}{\nano\meter}$) in 0.5x TBE buffer, as a function of the average diameter $D$ (geometric average of width and depth), in \SI{180}{\nano\meter} deep
nanochannels. The ``error bars'' denote the standard deviation of the measured extension at a given $D$.
} 
\label{Fig:0k5} 
\end{figure}

\begin{figure}
\includegraphics[width=12cm]{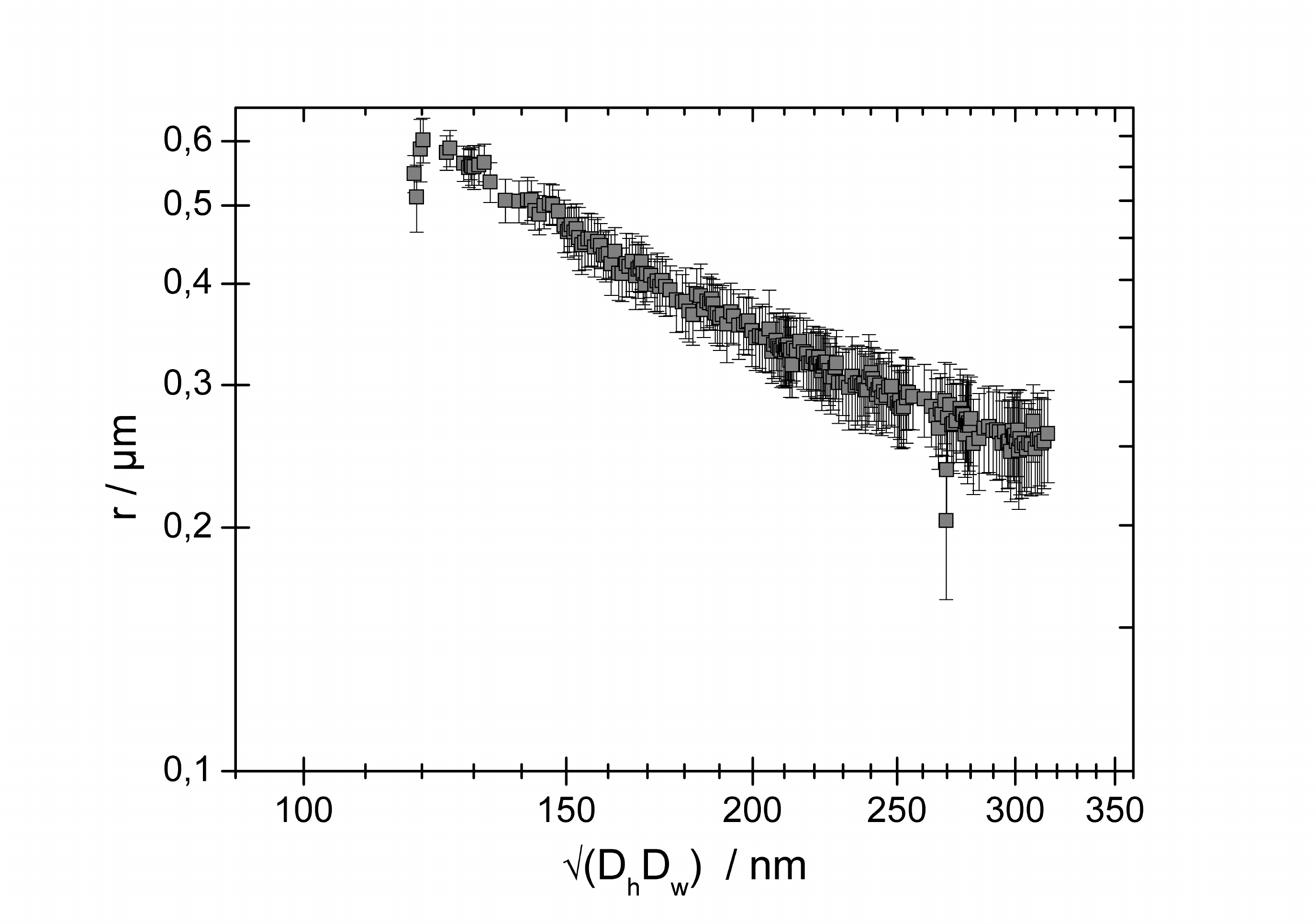}
\caption{The extension $r$ of $\lambda$-DNA (contour length $L\approx \SI{20}{\nano\meter}$) in 0.05x TBE buffer, as a function of the average diameter $D$ (geometric average of width and depth), in \SI{180}{\nano\meter} deep
nanochannels. The ``error bars'' denote the standard deviation of the measured extension at a given $D$.
} 
\label{Fig:0k05} 
\end{figure}

\noindent{\bf Measurement details.} All experiments were conducted with $\lambda$-phage DNA (48.5 kbp, New England Biolabs). The DNA was stained with the fluorescent dye YOYO\textsuperscript{\textregistered}-1 (Invitrogen) at a concentration of one dye molecule per ten base pairs. The DNA was dissolved in one of three three different buffers: 1x TBE buffer (\SI{89}{\milli M} tris(hydroxymethyl)aminomethane (TRIS), \SI{89}{\milli M} sodium borate, \SI{2}{\milli M} ethylenediaminetetraacetic acid (EDTA)) (\fig{1}), 0.5x TBE buffer (\fig{0k5}), or 0.05x TBE buffer (\fig{0k05}). In all cases, 3\% $\beta$\nobreakdash-mercaptoethanol was added to suppress photobleaching of YOYO-1. Note that this addition contributes to the ionic strength. Prior to every movie recording the DNA was left for 1 min in order to thermally equilibrate after being shifted along the channel by applying a pressure gradient along the nanochannel. To control that this equilibration time was sufficient, 5 movies were also recorded after 10 
min. We did 
not observe any significant change in the extension of the DNA compared to the movies recorded after 1 min of 
equilibration. Additionally, the relaxation time for $\lambda$-DNA in nanochannels was measured in \cite{reisner2005} and found to be on the order of \SI{1}{\second}, which indicates that 1 min of equilibration is enough.
Note that the gradient in confinement free energy was not sufficient to induce any detectable motion in the DNA during the experimentally relevant timescales.\\

\begin{figure}
\includegraphics[width=12cm]{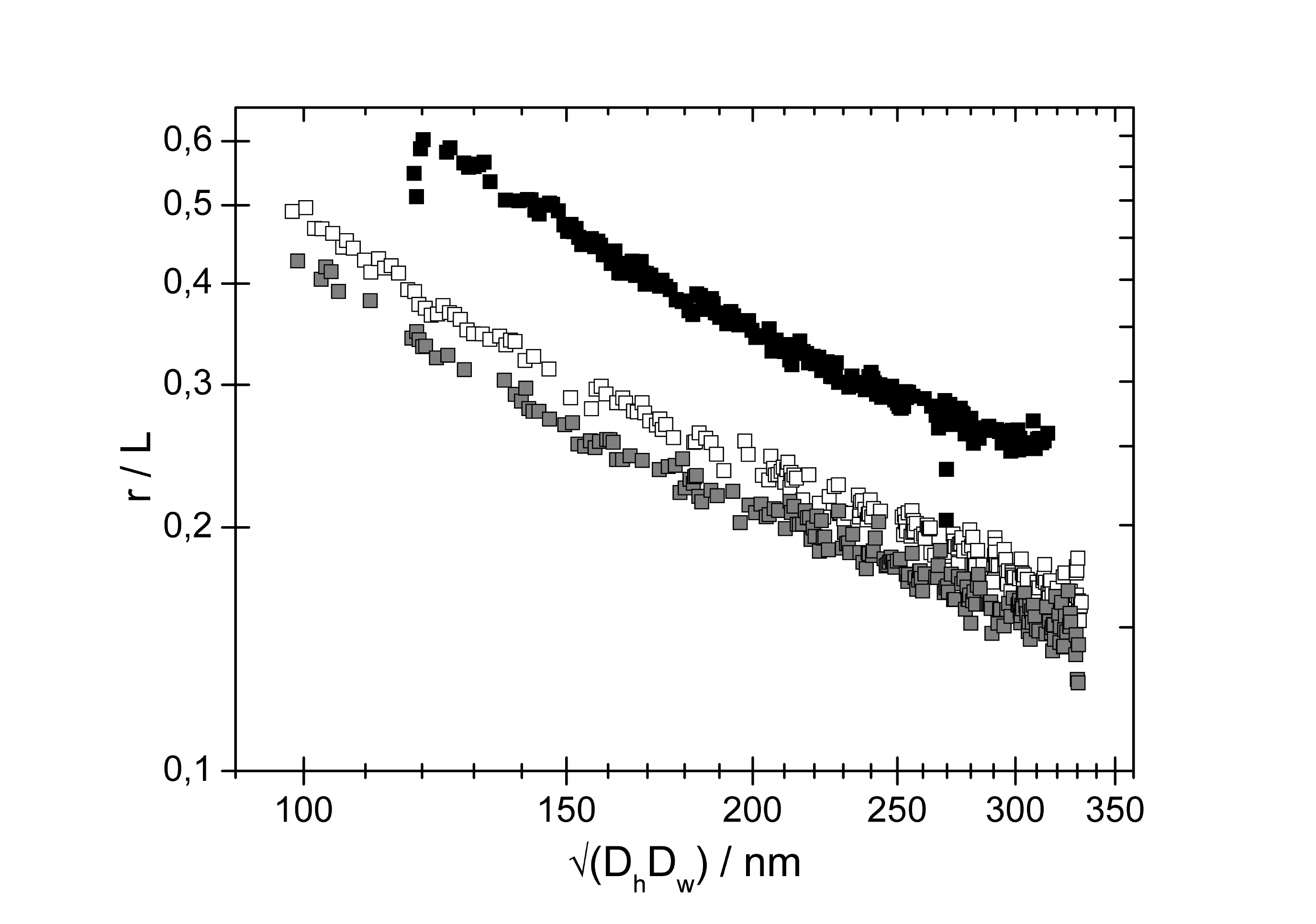}
\caption{The average extension $r$ of $\lambda$-DNA (contour length $L\approx \SI{20}{\nano\meter}$) as a function of the average diameter $D$ (geometric average of width and depth) in \SI{180}{\nano\meter} deep
nanochannels, studied at three different ionic strengths: 1x TBE (gray), 0.5x TBE (white), and 0.05x TBE (black).
} 
\label{Fig:All} 
\end{figure}

\noindent{\bf Data analysis.} After equilibration, the molecule was photographed for 400 frames, with a frame rate of 10 Hz. From each frame, we extracted the average diameter $D=\sqrt{D_{\rm h}D_{\rm w}}$ (defined as the geometric average of the width and depth of the channel at the center of the molecule) and the extension $r$. The resulting extensions from approximately 100 molecules were sorted by diameter, and collected into bins of 200 measurements each. For each of the bins, the average and the standard deviation of the extension was measured. The resulting data is shown in Fig.~\ref{Fig:All}\nobreakdash-\ref{Fig:0k05}.
For further details on the data analysis, and a description of how the extension $r$ was determined for each frame, we refer to \cite{PerssonReview2010}.\\

